\documentclass{elsart}

\usepackage{epsfig}
\usepackage{amssymb}

\begin{document}
\begin{frontmatter}

\title
{Olami-Feder-Christensen Model \\ on different Networks}

\author
{Filippo Caruso$^{1}$, Vito Latora$^{2}$, Alessandro Pluchino$^{2}$,}
\author
{Andrea Rapisarda$^{2}$ and Bosiljka Tadi\'c$^{3}$}

\address
{$^{1}$ Scuola Superiore di Catania, Via S. Paolo 73, I-95123
Catania,
Italy\\
$^{2}$Dipartimento di Fisica e Astronomia, Universit\`a di
Catania, \\ and INFN sezione di Catania, Via S. Sofia 64, I-95123
Catania, Italy\\
$^{3}$Department  for Theoretical Physics, Jo\v{z}ef Stefan
Institute, P.O. Box 3000; SI-1001 Ljubljana, Slovenia}

\maketitle
\begin{abstract}
We investigate numerically the Self Organized Criticality (SOC)
properties of the dissipative Olami-Feder-Christensen model on
small-world and scale-free networks. We find that the small-world
OFC model exhibits self-organized criticality. Indeed, in this
case we observe power law behavior of earthquakes size
distribution with finite size scaling for the cut-off region. In
the scale-free OFC model, instead, the strength of disorder
hinders synchronization and does not allow to reach a critical
state.
\end{abstract}

\end{frontmatter}

\section{Introduction}
The idea of the seismogenic crust as a self-organized complex
system was introduced over the years as a possible explanation for
the widespread occurrence of space-time long-range correlations in
earthquakes dynamics, similar to those observed in critical phase
transitions \cite{stanley}. In general, the term self-organized
criticality (SOC) \cite{BTW} refers to the intrinsic tendency of a
large class of spatially extended dynamical systems to
spontaneously organize into a dynamical critical state. One
signature of SOC is the presence of both a power law behavior in
earthquakes size distributions and a finite size scaling for their
cutoffs. Among the great number of different SOC models
\cite{bak_book,jen_book} developed in the last years, the OFC
model \cite{Olami}, introduced by Olami, Feder and Christensen in
1992, has played a key role in modelling earthquakes
phenomenology. However the presence of criticality in the
non-conservative version of this model has been controversial
since its introduction \cite{klein} and it is still
debated~\cite{carvalho,kim}, also in relation with the influence
of topology. In literature, OFC models on different topologies has
been investigated, in particular the 2D nearest neighbor lattice
(NNL) model \cite{lisepac1}, annealed random neighbor (ARN) graph
model and the OFC model on a quenched random (QR) graph
\cite{Lise}. The purpose of our work is to study the effects of
small-world (SW) and scale-free (SF) topologies on the criticality
of the non-conservative OFC model.
\\
The paper is organized in the following way. In section
\ref{OFCmodel} we review the OFC model and we point out the main
reasons that have induced us to study the non-conservative OFC
model on SW and SF topologies. In section \ref{SWnetwork} we
investigate the SW OFC model: in subsection \ref{SWsize} we show
the earthquakes size distributions for the non-conservative SW OFC
model and in subsection \ref{SWfss} we characterize the critical
behavior of the model through the finite size scaling ansatz.
Finally, in section \ref{SFmodel} we investigate the OFC model on
a scale-free network, obtained by preferential attachment
procedure \cite{albert}. Conclusions are drawn in section
\ref{Conclusions}.

\section{The Olami-Feder-Christensen model}
\label{OFCmodel} 
\begin{figure} [h]
\begin{center}
\includegraphics[width=9.5cm]{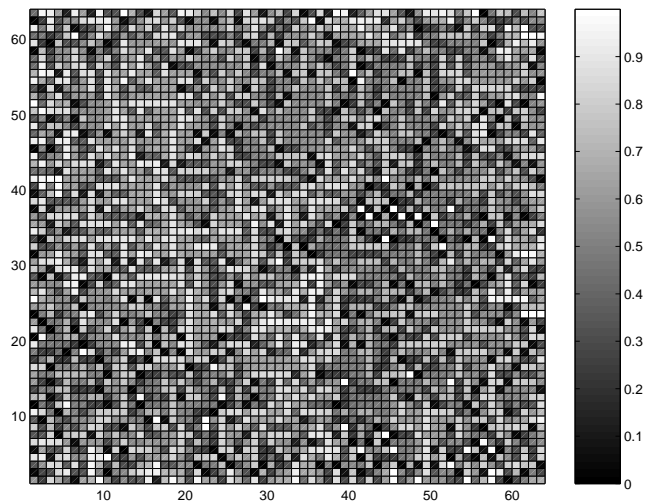}
\caption{\label{stress} Critical stress field of a 64x64 lattice
(NNL OFC model) in the critical state.}
\end{center}
\end{figure}

The Olami-Feder-Christensen (OFC) model
\cite{Olami} is defined on a discrete system of $N$ sites (blocks
or of fault elements) on a square lattice, each carrying a
seismogenic force (see Fig. \ref{stress}). Such a force is
simulated by associating to each site a real variable $F_i$, which
initially takes a random value in the interval $(0,F_{th})$.  All
the forces are increased simultaneously and uniformly (mimicking a
uniform tectonic loading), until one of them reaches the threshold
value $F_{th}$ and becomes unstable $(F_i \geq F_{th})$. The
uniform driving is then stopped and an ``earthquake'' (or
avalanche) starts:
\begin{equation}
\label{av_dyn}
F_i \geq F_{th}  \Rightarrow \left\{ \begin{array}{l}
                                       F_i \rightarrow 0 \\
                         F_{nn} \rightarrow F_{nn} + \alpha F_i
                                      \end{array} \right.
\end{equation}
where ``nn'' denotes the set of nearest-neighbor sites of $i$. The
parameter $\alpha$ controls the level of conservation of the
dynamics and, in the case of a graph with fixed connectivity $q$,
it takes values between $0$ and $1/q$ ($\alpha=1/q$ corresponding
to the conservative case). The toppling rule (\ref{av_dyn}) can
possibly create new unstable sites, producing a chain reaction.
All sites that are above threshold at a given time step in the
avalanche relax simultaneously according to (\ref{av_dyn}) and the
earthquake is over when there are no more unstable sites in the
system ($F_i < F_{th}$, $\forall i$). The uniform growth then
starts again. The number of topplings during an earthquake defines
its size, $s$, and we will be interested in the probability
distribution $P_N (s)$. In the following the boundary conditions
of the model will be ``open'', implying that $F=0$ on the boundary
sites.
\\
At this point it is important to emphasize that the OFC model behavior strongly depends
on the chosen topology.
For instance, in the dissipative NNL OFC model with open boundary conditions
the earthquakes size
distribution is described by a power law \cite{lisepac1,lisepac2}, characterized by a
universal exponent $\tau \simeq 1.8$ independent of the
dissipation parameter. However, at variance with the conservative case where
a full SOC behavior is observed, finite size scaling
appears to be violated in the pdf cutoff-region (see Table \ref{table1}).
%%%%%%%%%%%%%%%%%%%%%%%%  FIGURE %%%%%%%%%%%%%%%%%%%%%%%
%%%%%%%%%%%%%%%%%%%%%%%%%%%%%%%%%%%%%%%%%%%%%%%%%%%%%%%%%%
\\
In ARN OFC models \cite{lise,chabanol,broker,kinouchi}, where each site
interacts with randomly chosen sites instead of its nearest
neighbors on the lattice, there is criticality only in the conservative case, where it
becomes equivalent to a critical branching process. As soon as
some dissipation is introduced, the earthquakes become localized
although the mean earthquakes size diverges exponentially as
dissipation tends to zero and there is no power law distribution (see Table \ref{table1}).
Actually it is interesting to point out that
criticality in the OFC model on a lattice has been
ascribed to a mechanism of partial
synchronization~\cite{middleton}.
In general the system shows a tendency to
self-organize into a periodic state~\cite{middleton,socolar,grass2} which
is frustrated by the presence of inhomogeneities such as the
boundaries. In addition, inhomogeneities induce partial
synchronization of the elements of the system building up long
range spatial correlations and a critical state is obtained. The
mechanism of synchronization requires an underlying spatial
structure and therefore cannot operate in an ARN model,
where each site is assigned new random neighbors at each update.
\\
In the OFC model on a QR graph, where the choice of neighbors is not annealed
but quenched and all the sites have exactly the same number of
nearest neighbors $q$ (both for $q=4$ and $q=6$), the dynamics
organizes into a subcritical state.
This is analogous to what happens in the OFC model on a NN lattice
with periodic boundary conditions, where no critical behavior is
observed at all \cite{middleton,socolar,grass2}. In the QR case,
in order to observe scaling in the earthquakes distribution, one has to introduce some
inhomogeneities \cite{Lise}. In particular,
it has been found that it is enough to consider just two sites in the
system with coordination $q-1$~\cite{Lise}. When either of these
sites topple according to rule (\ref{av_dyn}), an extra amount $
\alpha F_i$ is simply lost by the system. In such a way spatial correlations can develop,
because the topology is quenched, there is power law in earthquakes size distribution and also finite size scaling
is observed (see Table \ref{table1}).
\\
\\
In this work we study the non-conservative OFC model on both a
small-world and a scale-free topology.
\\
First of all we expect that the inclusion of some inhomogeneities
in the sites degree is not the unique way to obtain SOC. Indeed,
as we are going to show, an alternative way is to keep fixed the
sites degree and to change the topology of the underlying network,
for instance by considering a small-world graph obtained by
randomizing a fraction $p$ of the links of the regular NN lattice.
Here we will use the term ``small-world'' to refer to a rewired
lattice (with fixed connectivity) with the minimum number of
rewired links such that the characteristic path length $L$ is
almost as small as that one for the corresponding random graph
\cite{watts,sw}. As shown in the right panel of Fig.
\ref{rewiring}, this is obtained already for very small values of
$p$ ($p \simeq 0.01$), much before the random graph limit ($p=1$).
\\
A small-world topology is expected to be a more accurate
description of a real system according to the most recent
geophysical observations that indicate that earthquakes
correlation might extend to the long range in both time and space
\cite{stein,tosi}. In fact, if a main fracture episode occurs, it
may induce slow strain redistribution through the earth crust,
thus triggering long-range as well as short-range seismic effects
\cite{kagan,hill,cresce,parsons,palatella}. The presence of a
certain percentage of long-range connections in the network takes
into account the possibility that an earthquake can trigger other
earthquakes not only locally but also at long distances.
\\
On the other hand, one can consider a different kind of networks
with a small $L$, the so called ``scale-free'' networks, which
differ from the small-world graphs for having a power law
distribution of the site degree. Scale-free networks are very
common in nature and have also been used for SOC models (see ref.
\cite{goh} for sandpile dynamics on SF network) but they have not
been investigated, as far as we know, in the context of OFC
models. It is known that, when the connectivity is not fixed at
all but only in average (as for a random graph in Ref.
\cite{Lise}), the strength of disorder is enough to destroy
critical behavior. Thus we expect that for SF networks, where the
connectivity has a power law distribution, synchronization will
not take place and it will be not possible for the OFC model to
reach a critical state. In the last part of the paper we will show
that this is exactly what happens.

\section{The OFC model on small-world network}
\label{SWnetwork} 
\begin{figure} [h]
\begin{center}
\includegraphics[width=8cm]{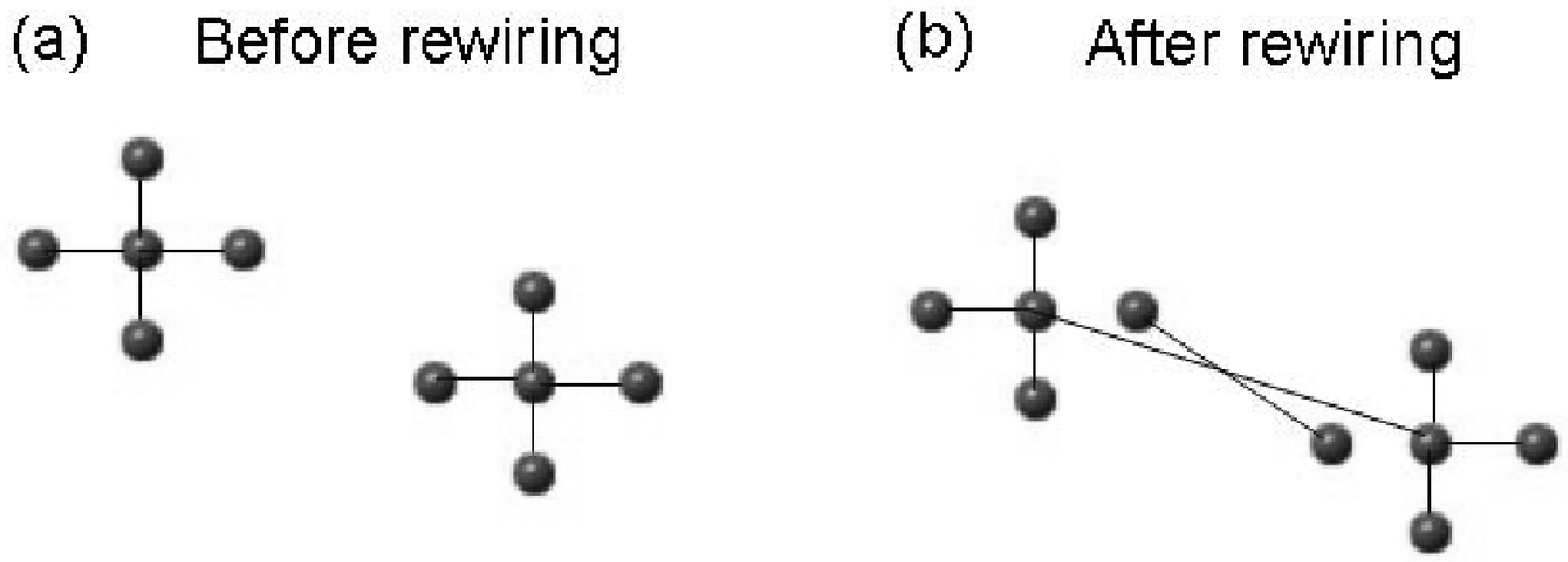}
\includegraphics[width=5cm]{CarusoFig2b}
\caption{\label{rewiring} On the left, a schematic picture of the
rewiring procedure to interpolate between a regular and a random
topology by keeping fixed and equal to $4$ the degree of each
site. On the right, we report the characteristic path length L vs.
the rewiring probability $p$.}
\end{center}
\end{figure}

To investigate the effects of the small-world
topology on the criticality of the non-conservative OFC model, we
follow the method proposed by Watts and Strogatz to construct a
network which interpolates between a square NN lattice and a
quenched random graph \cite{watts,wattsbook}. We start with a
two-dimensional NN square lattice in which each site is connected
to its $4$ nearest neighbors. The links of the lattice are rewired
at random with a probability $p$ as in the one-dimensional model
of Ref. \cite{watts}. The main differences with respect to the
original model is that for any value of $p$ we want to keep fixed
the connectivity of each site. For such a reason we have
implemented a rewiring procedure as in fig.\ref{rewiring} in which
the connections are rewired in couples.
%%%%%%%%%%%%%%%%%%%%%%%%  FIGURE %%%%%%%%%%%%%%%%%%%%%%%
%%%%%%%%%%%%%%%%%%%%%%%%%%%%%%%%%%%%%%%%%%%%%%%%%%%%%%%%%
We choose a site $i_1$ and the edge $i_1-i_2$ that connects site
$i_1$ to its nearest neighbor $i_2$ in a clockwise sense. With
probability $p$ we decide whether to rewire this edge or to leave
it in place. If the edge has to be rewired we (a)
 choose at random a second site $j_1$ and one of its edges,
for instance the edge $j_1-j_2$ connecting site $j_1$ to site
$j_2$, and (b) we substitute the couple of edges $i_1-i_2$ and
$j_1-j_2$ with the couple $i_1-j_2$ and $j_1-i_2$.
\\
We repeat this process by moving over the entire square lattice
considering each site in turn until one lap is completed. In such
a way the limit case $p=1$ is a QR graph with fixed connectivity
(q) equal to $4$. In the intermediate cases $0<p<1$ we can
investigate the effects of an increasing number of long-range
connections on the criticality of the model. Indeed, at a critical
region of the parameter $p$ between the regular ($p=0$) and random
($p=1$) networks, the topology produced by such a method exhibits
a small-world behavior, characterized by the fact that the
distance between any two sites on the graph is of the order of
that for a random network and, at the same time, the concept of
neighborhood is preserved, as for regular lattices. For this
reason, we expect to obtain SOC in a small-world topology; the
introduction of a few long-range edges create short-cuts that
connect sites that otherwise would be much further apart.

\subsection{Earthquakes size distributions}
\label{SWsize}
\begin{figure} [h]
\begin{center}
\includegraphics[width=9.5cm, angle=0]{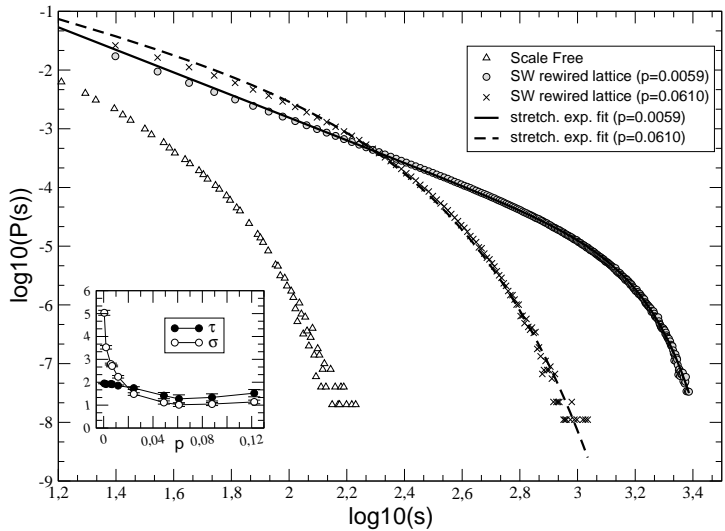}
\caption{\label{stretched} Earthquakes size distributions for the
non-conservative OFC model (with $\alpha=0.21$) on rewired 64x64
lattice with two rappresentative values of rewiring probability
$p$ in the range [0,0.1]. A stretched-exponential function is used
to fit the pdf's cutoffs. In the inset we plot the two exponents
$\tau$ and $\sigma$ as a function of the rewiring probability $p$.
Here we report also the pdf for SF OFC model (see sec.
\ref{SFmodel}).}
\end{center}
\end{figure}

In our simulations the starting point for the
construction of the SW network is a two-dimensional square lattice
$L \times L$ with three different sizes: $L=32, 64$ and $128$; the
corresponding number of sites is $N= L^2$. We have considered up
to $10^9$ earthquakes to obtain a good statistics for the
earthquakes size distribution $P_N(s)$. In fig.\ref{stretched} we
report the power law distributions resulting for $N=64^2$,
$\alpha=0.21$ (non-conservative OFC model) and for two
rappresentative values of the rewiring probability $p$ (actually
we made the simulations also for many other values of $p$ in the
range [0,0.1]). In the same figure we report also the comparison
with the earthquakes size distribution for the dissipative OFC
model on a scale-free network (that will be discussed in section
\ref{SFmodel}).
%%%%%%%%%%%%%%%%%%%%%%%%  FIGURE %%%%%%%%%%%%%%%%%%%%%%%
%%%%%%%%%%%%%%%%%%%%%%%%%%%%%%%%%%%%%%%%%%%%%%%%%%%%%%%%
\\
All the curves can be fitted by a stretched-exponential function
$P_N(s)=A s^{-\tau} e^{-(s/\xi)^\sigma}$,
where $s$ is the size of earthquakes, $\xi$ is the characteristic
length and $\tau$ and $\sigma$ are two exponents. We notice that,
increasing more and more the rewiring probability, the power law is practically
lost. This can
be better exploited by plotting the value of the two exponents
$\tau$ and $\sigma$ as a function of $p$ in the inset in Fig.\ref{stretched}.
Indeed one can expect stretching-exponential in various cases of
stochastic processes where many length scales appear. We note also that, above the value
$p \simeq 0.01$ for which $\sigma$ suddenly approaches $1$, the power
law for the pdfs progressively disappears. In the next subsection we will
show that the cut-off in the earthquakes probability distribution scales
with the system size (the so called finite size scaling ansatz) only around this rewiring threshold.

\subsection{Finite Size Scaling}
\label{SWfss} 
\begin{figure} [h]
\begin{center}
\includegraphics[width=7cm,angle=0]{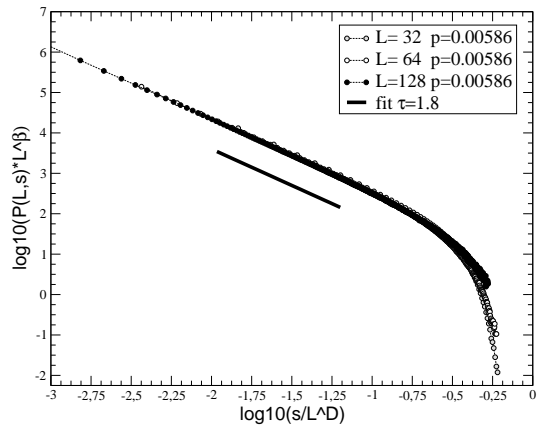}
\bigskip
\caption{\label{fss} Finite Size Scaling for dissipative OFC model
(with $\alpha=0.21$) on a small-world topology for three different
values of $N$, namely $N=32^2, 64^2, 128^2$. The critical exponent
derived from the fit are $D=2$ and $\beta \simeq 3.6$ and the
rewiring probability is equal to $0.00586$.}
\end{center}
\end{figure}

In order to characterize the critical behavior of
the dissipative SW OFC model, a finite size scaling (FSS) ansatz
is applied, i.e. $P_N(s) \simeq N^{-\beta} f(s/N^D)$ where $f$ is
a suitable scaling function and $\beta$ and $D$ are critical
exponents describing the scaling of the distribution function. In
Fig. \ref{fss} we consider $\alpha=0.21$ and a rewiring
probability $p \simeq 0.006$. We show the collapse of $P_N(s)$ for
three different values of $N$, namely $N=32^2, 64^2, 128^2$. The
distribution $P_N(s)$ satisfies the FSS hypothesis reasonably
well, with universal critical coefficients with small rewiring
probability, but, increasing $p$, as shown in the previous
subsection, there is no power law at all. The critical exponent
derived from the fit of Fig. \ref{fss} are $\beta \simeq 3.6$ and
$D=2$. This result is in agreement with the FSS hypothesis
implying that, for asymptotically large $N$, $P_N(s) \sim
s^{-\tau}$ with $\tau= \beta/D \simeq 1.8$.
\\
Therefore, showing both power law behavior and FSS, the dissipative SW OFC model (in a restricted
range of rewiring probability) clearly exhibits self-organized criticality (see Table \ref{table1}).
Let us point out that on the SW rewired topology the system behaves as in the compact square
lattice, but the occurrence of a small amount of long-range links
disseminate the earthquakes over the network and the biggest
earthquake size scales with the lattice. On the other hand,
if we increase the amount of long-range links above a certain threshold
($p \simeq 0.006$), the mechanism of synchronization is corrupted and
the scaling behavior disappears.

\section{The OFC model on a scale-free network}
\label{SFmodel}
Finally we investigate criticality of the non-conservative OFC
model on a scale-free network. It is an example of network displaying
a small characteristic path length and a power-law distribution $p(k)\sim
k^{-\gamma}$ in the node connectivity k (degree). By using the preferential attachment growing procedure
introduced by Barab\'{a}si and Albert \cite{albert}, we start from $m+1$ all to all
connected nodes and at each time step we add a new node with $m$
links. These $m$ links point to old nodes with probability
$p_i=\frac{q_i}{\sum_j q_j}$, where $q_i$ is the degree of the
node $i$. This procedure allows a selection of the $\gamma$
exponent of the power law scaling in the degree distribution with
$\gamma=3$ in the thermodynamic limit ($N \longrightarrow
\infty$). Here we consider a scale-free network with $\gamma=3$ and $N=1000$.
\\
In this case, the toppling rule in Eqs. \ref{av_dyn} must be
modified to take into account that different sites have a
different coordination number $q_{i}$. Each site consequently has
a different $\alpha_i$, which we determined by requiring that the
total fraction $\tilde{\alpha}$ of the force transferred from the
unstable site to the nearest-neighbor sites is constant in the
system, i.e., $\alpha_i=\tilde{\alpha}/q_i$; here we consider the case $\tilde{\alpha}=0.21$.
\\
We have found that there is no criticality in the system since
there is no power law in the earthquakes size distribution, as
shown in fig. \ref{stretched}. As previously observed and in
agreement with previous investigations
\cite{lisepac1,ceva,mousseau}, this result indicates that if the
disorder is too strong the critical signatures disappear and the
SOC behavior is destroyed (see Table \ref{table1}).

\begin{table*}
\begin{center}
\begin{tabular}{ccccccc}
\hline
 & & \textbf{TOPOLOGY} & & \textbf{Power Law} & & \textbf{Finite Size Scaling}\\ \hline
 & &\textbf{2D NN lattice} & & Yes & & No\\
 & &\textbf{ARN graph} & & No & & No\\
 & &\textbf{QR graph} & & No & & No\\
 & &\textbf{QR graph+2} & & Yes & & Yes\\
 & &\textbf{SW network} & & Yes & & Yes\\
 & &\textbf{SF network} & & No & & No\\ \hline
\end{tabular}
\end{center}
\bigskip
\caption{In this table we list the SOC properties for OFC models
on different topologies: a two dimensional nearest neighbor
lattice (2D NN lattice), an annealed random neighbor (ARN) graph,
a quenched random (QR) graph, a quenched random (QR+2) graph with
two sites with coordination 3, a small-world (SW) network and a
scale-free (SF) network. We always consider these models in the
case $\alpha=0.21$ (dissipative regime) and with open boundary
conditions. In particular, we report when there is power law and
finite size scaling in earthquakes size distribution, according to
each kind of topology.} \label{table1}
\end{table*}

\section{Conclusions}
\label{Conclusions}
In conclusion, in this paper we have investigated the dissipative OFC model on
small-world and scale-free networks.
\\
We have shown that, at variance with
OFC models on other topologies which are critical
only in the conservative case, the dissipative small-world OFC model clearly reaches a critical state characterized
by power law behavior of earthquakes size distribution
with finite size scaling of cut-offs.
Indeed, in a lattice with a small number
of rewired links the underlying spatial structure allows partial synchronization of
distant blocks of the system. We think that this process could reproduce the long-range
earthquakes dynamical correlations in the earth crust, according to the most recent geophysical observations.
\\
On the other hand, on a scale-free topology we do not observe SOC
properties. We expected this behavior because the connectivity is
not fixed; so the dynamics is not synchronized, the disorder is
too strong and the critical state is destroyed. As future
directions, it seems interesting to better investigate the
influence of topology and the role of disorder on the
self-organized criticality properties of the OFC models.


\begin{thebibliography}{99}

\bibitem{stanley}
E. Stanley, {\it Introduction to Phase Transitions and Critical
Phenomena}, (Oxford University Press, 1987).

\bibitem{BTW}
P. Bak, C. Tang, and K. Wiesenfeld,
Phys. Rev. Lett. {\bf59}, 381 (1987);
Phys. Rev. A. {\bf 38}, 364 (1988).

\bibitem{bak_book}
P. Bak, {\it How Nature Works: The Science of Self-Organized Criticality}
(Copernicus, New York, 1996).

\bibitem{jen_book}
H. Jensen,
 {\it Self-Organized Criticality}
      (Cambridge Univ. Press,New York, 1998).

\bibitem{Olami}
Z. Olami, H.J.S. Feder, and K. Christensen, Phys. Rev. Lett. {\bf
68}, 1244 (1992);
 K. Christensen and Z. Olami,
Phys. Rev. A {\bf 46}, 1829 (1992).

\bibitem{klein}
W. Klein and J. Rundle, Phys. Rev. Lett. {\bf 71}, 1288 (1993);
 K. Christensen, Phys. Rev. Lett. {\bf 71}, 1289 (1993).

\bibitem{carvalho}
J.X. Carvalho and C.P.C. Prado, Phys. Rev. Lett. {\bf 84}, 4006
(2000).

\bibitem{kim}
K. Christensen, D. Hamon, H.J. Jensen, and S. Lise, Phys. Rev.
Lett. {\bf 87}, 039801 (2001); J.X. Carvalho and C.P.C. Prado,
Phys. Rev. Lett. {\bf 87}, 039802 (2001).

\bibitem{lisepac1}
S. Lise and M.Paczuski, Phys. Rev. E, {\bf 63}, 036111 (2001).

\bibitem{Lise} S. Lise and M. Paczuski, Phys. Rev. Lett. 88, 228301 (2002).

\bibitem{albert}
A. L. Barab\'{a}si and R. Albert, Science {286}, 509 (1999).

\bibitem{lisepac2} S. Lise and M.Paczuski, Phys. Rev. E,
{\bf 64}, 046111 (2001).

\bibitem{lise} S. Lise and H. J. Jensen, Phys. Rev. Lett. {\bf
76}, 2326 (1996).

\bibitem{chabanol}
M. L. Chabanol and V. Hakim, Phys. Rev. E {\bf 56}, 2343 (1997).

\bibitem{broker}
H. M. Broker and P. Grassberger, Phys. Rev. E {\bf 56}, 3944
(1997).

\bibitem{kinouchi}
O. Kinouchi, S.T.R. Pinho, and C.P.C. Prado, Phys. Rev. E {\bf
58}, 3997 (1998).

\bibitem{middleton}
A. A. Middleton and C. Tang, Phys. Rev. Lett. {\bf 74}, 742
(1995).

\bibitem{socolar}
J.E.S. Socolar, G. Grinstein, and C. Jayaprakash, Phys. Rev E,
{\bf 47}, 2366 (1993).

\bibitem{grass2}
P. Grassberger, Phys. Rev. E {\bf 49}, 2436 (1994).

\bibitem{watts}
D.J. Watts and S.H. Strogatz,
Nature \textbf{393}, 440 (1998).

\bibitem{sw}
More in general in Wattz and Strogatz original definition a
small-world network is characterized not only by a small value of
$L$, but also by an high clustering coefficient.

\bibitem{stein}
R. S. Stein, Nature \textbf{402}, 605 (1999).

\bibitem{tosi}
P. Tosi, V. De Rubeis, V. Loreto, and L. Pietronero,
Annals of Geophysics,\textbf{47}, 1849 (2004).

\bibitem{kagan}
Y.Y. Kagan and D.D. Jackson, Geophys. J. Int. 104, 117 (1991).

\bibitem{hill}
D.P. Hill et al., Science 260, 1617 (1993).

\bibitem{cresce}
L. Crescentini, A. Amoruso, R. Scarpa,
Science {286}, 2132 (1999).

\bibitem{parsons}
T. Parsons, J. Geophys. Res., 107, 2199 (2001).

\bibitem{palatella}
M. S. Mega, P. Allegrini, P. Grigolini,
V. Latora, L. Palatella, A. Rapisarda
and S. Vinciguerra, Phys. Rev. Lett. {\bf 90}, 188501 (2003).

\bibitem{goh}
K.-I. Goh, D.-S. Lee, B. Kahng, and D. Kim,
Phys. Rev. Lett. {\bf 91}, 148701 (2003).

\bibitem{wattsbook}
D.J. Watts \textit{Small Worlds} (Princeton Univ. Press, Princeton,
New Jersey, 1999).

\bibitem{ceva}
H. Ceva, Phys. Rev. E {52}, 154 (1995).

\bibitem{mousseau}
M. Mousseau, Phys. Rev. Lett. {77}, 968 (1996).

\end{thebibliography}
\end {document}